\documentclass[12pt]{article}
\usepackage{epsfig}
\usepackage{latexsym}
\usepackage{color}
\newcommand{\mysquare}[0]{\raise-.2ex\hbox{{\Large$\Box$}}}
\def\lsim{\mathrel{\rlap {\raise.5ex\hbox{$ < $}}
{\lower.5ex\hbox{$\sim$}}}}
\def\gsim{\mathrel{\rlap {\raise.5ex\hbox{$ > $}}
{\lower.5ex\hbox{$\sim$}}}} \topmargin -1.5cm \textheight=22.5cm
\catcode`\@=11
\newcount\hour
\newcount\minute
\newtoks\amorpm
\hour=\time\divide\hour by60
\minute=\time{\multiply\hour by60 \global\advance\minute by-\hour}
\edef\standardtime{{\ifnum\hour<12 \global\amorpm={am}%
        \else\global\amorpm={pm}\advance\hour by-12 \fi
        \ifnum\hour=0 \hour=12 \fi
        \number\hour:\ifnum\minute<10 0\fi\number\minute\the\amorpm}}
\edef\militarytime{\number\hour:\ifnum\minute<10 0\fi\number\minute}
\def\draftlabel#1{{\@bsphack\if@filesw {\let\thepage\relax
   \xdef\@gtempa{\write\@auxout{\string
      \newlabel{#1}{{\@currentlabel}{\thepage}}}}}\@gtempa
   \if@nobreak \ifvmode\nobreak\fi\fi\fi\@esphack}
        \gdef\@eqnlabel{#1}}
\def\@eqnlabel{}
\def\@vacuum{}
\def\draftmarginnote#1{\marginpar{\raggedright\scriptsize\tt#1}}
\def\draft{\oddsidemargin -.2truein
        \def\@oddfoot{\sl preliminary draft \hfil
        \rm\thepage\hfil\sl\today\quad\militarytime}
        \let\@evenfoot\@oddfoot \overfullrule 3pt
        \let\label=\draftlabel
        \let\marginnote=\draftmarginnote
   \def\@eqnnum{(\theequation)\rlap{\kern\marginparsep\tt\@eqnlabel}%
\global\let\@eqnlabel\@vacuum}  }

\relax

\newcommand{\ba}[0]{\begin{eqnarray}}
\newcommand{\ea}[0]{\end{eqnarray}}
\def\bs{\begin{subequations}}
\def\es{\end{subequations}}

\def\np#1#2#3{Nucl. Phys. {\bf{B#1}} (#2) #3}
\def\pl#1#2#3{Phys. Lett. {\bf{B#1}} (#2) #3}

\def\thebibliography#1{%
\vskip 0.5cm \centerline{\bf References}
\list{%
[\arabic{enumi}]}{\settowidth\labelwidth{[#1]}
\leftmargin\labelwidth
\advance\leftmargin\labelsep
\usecounter{enumi}}
\def\newblock{\hskip .11em plus .33em minus .07em}
\sloppy\clubpenalty4000\widowpenalty4000
\sfcode`\.=1000\relax}

\renewcommand{\theequation}{\arabic{section}.\arabic{equation}}

\renewcommand{\section}{\setcounter{equation}{0}\@startsection%
{section}{1}{0mm}{-\baselineskip}{0.5\baselineskip}%
{\normalfont\normalsize\bfseries}}

\renewcommand{\subsection}{\@startsection%
{subsection}{2}{0mm}{-\baselineskip}{0.5\baselineskip}%
{\normalfont\normalsize\slshape}}

\usepackage{amscd} \usepackage{amsmath} \usepackage{amsfonts}

\def\thefootnote{\fnsymbol{footnote}}
\def\es{\end{subequations}}

\def\e{\, {\rm  e }}

\def\nn{\nonumber}

\newcommand{\uarrw}[0]{\mathrel{
{\raise.5ex\vbox{\hrule width 1cm}\hskip-6pt\rightarrow}}}
\begin{document}
\renewcommand{\theequation}{\arabic{section}.\arabic{equation}}
\begin{titlepage}
\begin{flushright}
LPTENS/03/29\\
CPTH-PC044.0803\\
hep-th/0312300 \\
\end{flushright}
\begin{centering}
\vspace{15pt}
{\bf FIVE-BRANE CONFIGURATIONS, CONFORMAL FIELD THEORIES}\\
\vspace{5pt}
{\bf AND THE STRONG-COUPLING PROBLEM} $^\ast$\\
\vspace{15pt} {\bf E. Kiritsis$^{\ 1,2}$, C. Kounnas$^{\ 3}$, P.M.
Petropoulos$^{\ 2}$
and J. Rizos$^{\ 3,4}$}\\
\vspace{7pt}
{\it $^1 $ Department of Physics, University of Crete, and FO.R.T.H.}\\
{\it P.O. Box 2208, 71003 Heraklion, GREECE}\\
\vspace{5pt} {\it $^2 $ Centre de Physique Th{\'e}orique, Ecole
Polytechnique $^\dagger
$}\\
{\it 91128 Palaiseau Cedex, FRANCE}\\
\vspace{5pt} {\it $^3 $ Laboratoire de Physique Th{\'e}orique de
l'Ecole Normale Sup{\'e}rieure $^\diamond$}\\
{\it 24 rue Lhomond, 75231 Paris Cedex, FRANCE}\\
\vspace{5pt}
{\it $^4 $ Physics Department, University of Ioannina}\\
{\it 45110 Ioannina, GREECE} \vspace{15pt}

{\bf Abstract}\\
\end{centering}
\vspace{5pt} Decoupling limits of physical interest occur in
regions of space--time where the string coupling diverges. This is
illustrated in the celebrated example of five-branes. There are
several ways to overcome this strong-coupling problem. We review
those which are somehow related to two-dimensional conformal field
theories. One method consists of distributing the branes over
transverse space, either on a circle or over a sphere. Those
distributions are connected to conformal field theories by
T-dualities or lead to a new kind of sigma model where the target
space is a patchwork of pieces of exact conformal-field-theory
target spaces. An alternative method we discuss is the
introduction of diluted F-strings, which trigger a marginal
deformation of an AdS$_3\times S^3\times T^4$ background with a
finite string coupling. Our discussion raises the question of
finding brane configurations, their spectrum, their geometry, and
their interpretation in terms of two-dimensional conformal models.

\noindent\textsl{Based on talks given by some authors at the Corfu
School and Workshops on High-Energy Physics, Greece, September 1
-- 20 2001, at the annual meeting of the Hellenic Society for the
Study of High-Energy Physics, University of Patras, Greece, April
25 -- 28 2002, and at the Second Crete Regional Meeting in String
Theory, Kolymbari, Greece, June 19 -- 28 2003.}
\begin{flushleft}
LPTENS/03/29\\
CPTH-PC044.0803\\
December 2003
\end{flushleft}
\hrule width 6.7cm \vskip.1mm{\small \small \small
$^\ast$\  Research partially supported by the EEC under the contracts
HPRN-CT-2000-00122, HPRN-CT-2000-00131, HPRN-CT-2000-00148 and HPMF-CT-2002-01898.\\
$^\dagger$\ Unit{\'e} mixte  du CNRS et de  l'Ecole Polytechnique,
UMR 7644.\\
$^\diamond$\  Unit{\'e} mixte  du CNRS et de l'Ecole Normale
Sup{\'e}rieure, UMR 8549.}
\end{titlepage}
\newpage
\setcounter{footnote}{0}
\renewcommand{\thefootnote}{\arabic{footnote}}

\setcounter{section}{0}
\section{Some motivations and ideas}

String theory unifies gauge interactions and gravity. Until
recently, it was developed along two main streams of quite distant
goals. On the one hand, string phenomenology aimed at recovering
the standard model and other supersymmetric low-energy theories.
On the other hand, string gravity was focusing on the search for a
variety of backgrounds that could help in understanding cosmology
and probe gravity at short distances (study quantum effects,
singularities, black-hole thermodynamics \dots).

The discovery of branes has deeply modified the previous landscape
of interests in a variety of ways. Fundamental constituents are
now BPS objects like F1-, NS5-, or D$p$-branes, and the
understanding of string theory goes through the investigation of
these objects, in all possible ways \cite{CHS, polchinski}.

A prime conceptual achievement of the above discovery was to
capture the string spectrum beyond perturbation theory. In turn
this allowed for the unification of various string vacua through
dualities, and led to M-theory that has brought its own
fundamental objects: the M2- and M5-branes \cite{witten}.

All these extended objects have introduced new insights and
techniques. In this framework, phenomenology and gravity
motivations and methods have become very close. The safest tool
for string phenomenology is {\it exact conformal field theory
(CFT),} since it guarantees an absolute handling over the
(perturbative) string spectrum. This includes orbifold models,
fermionic constructions \dots For gravity purposes these methods
are not convenient because they often lack of clear geometrical
interpretation which is, however, straightforward in {\it
$O(\alpha ')$ two-dimensional sigma-models.} Those turn out to be
more useful for cosmology searches.

The presence of branes, blurs the latter Cartesian picture.
Indeed, branes act like impurities, which {\it alter the string
spectrum and modify the background}. A plethora of possibilities
for phenomenology with an up-to-date point of view on low
supersymmetry-breaking scale, hierarchies \dots, as well as new
geometrical set-ups with non-trivial gravitational (and other
background) fields therefore appear. Hence, a fundamental and
unifying aim emerges: {\it find brane configurations, their
spectrum, their geometry, and CFT interpretation}
\cite{hw,maldacena,w,l,a}. The Randall--Sundrum model is a good
illustration of this line of thought: it provides a brane
configuration with both geometrical and phenomenological intrinsic
value.

Another important drawback of the study of branes was the
discovery of \textit{decoupling limits different from the usual
low-energy limits} \cite {Seiberg:1997zk, Aharony:1998ub,
Itzhaki:1998dd}. The existence of such limits is quite unexpected
for simple reasons. String spectra are highly constrained (GSO
projections, modular invariance, supersymmetry issues \dots).
Gravity and gauge sectors appear therefore in an intricate way. It
looks very unlikely that one could find a limit -- other than
going to low energies -- which would enable us to disentangle this
imbroglio of states and separate those sectors; tracing back their
respective origins to some geometric feature looks even more out
of reach. Much like orbifold fixed points, branes contribute part
of the spectrum, though they truly appear to carry it in the
semi-classical limit only. Put differently, not only many sectors
must be considered to create a string, but these sectors cannot be
chosen at wish or designated to originate from a particular
geometrical object. Any ad hoc construction of this type is not
expected to survive -- as a string -- $O(\alpha')$ corrections.
The Randall--Sundrum model is again a good example of this
situation since no reliable string realization has been provided
so far, which can reproduce its features. The role of brane
configurations which possess a clear CFT interpretation appears
again to be of major importance.

Despite this highly constrained structure, there are limits such
that the spectrum (\romannumeral1) is dominated by excitations
leaving on the branes, (\romannumeral2) cannot be described by
means of ordinary low-energy quantum field theory, and
(\romannumeral3) does not contain the gravitational sector. This
decoupling of gravity is paradoxically an asset. The reason for
that is that, under these circumstances, old ideas about
\textit{holography} become operational.

Unfortunately, these decoupling limits occur in regions of
space--time where the string coupling diverges. Although these
divergences are physical (they are even intimately related to the
very existence of the decoupling) they set the limits to the
confidence of perturbative string theory. Alternative brane
configurations or regulated descriptions must be found, which can
be analyzed as much as possible, by means of exact CFT.

The motivation of this short note is to illustrate these issues in
the celebrated example of five-branes, and review various ways to
overcome the strong-coupling problem within frameworks that are
related, one way or the other, to two-dimensional conformal
models.

\section{The five-brane solutions in type II superstring}
\label{5brem}

The ten-dimensional effective action reads, in the Einstein frame:
\begin{equation}
S^{(10)}=\frac{1}{2\kappa^2_{10}}\int d^{10}x \,
\sqrt{-g^{(10)}}\left(
R^{(10)}-\frac{1}{2}(\partial\phi_\gamma)^2-\frac{1}{12}
\e^{-\gamma\phi_\gamma}\,H_{\gamma}^2\right). \label{act}
\end{equation}
Here $\phi_\gamma$ is the dilaton field and $\gamma=\pm 1$
corresponds to the two distinct NS--NS or R--R three-form field
strengths $H$ in type IIB theory (type IIA allows only for
$\gamma=+1$). We do not introduce any gauge field, which means in
particular that the branes under consideration carry no other
charge than NS--NS or R--R.

The canonical five-brane solutions are of the form
\begin{equation}
ds^2=h(r)^{-1/4}\left(-dt^2+d\vec{x}^2\right)+ h(r)^{3/4}
\left(dr^2 + r^2 d\Omega^2_3\right), \label{5met}
\end{equation}
where $\vec{x}\equiv \{x^5,x^6,\ldots,x^9 \}$ are Cartesian
coordinates in a five-dimensional Euclidean flat space and $
d\Omega^2_3$ is the metric on a unit-radius three-sphere. Together
with the radial (dimensionless) coordinate $r$, the latter is
transverse to the five-brane. Poincar{\'e} invariance within the
five-brane world-volume is here automatically implemented.

The ansatz (\ref{5met}) indeed minimizes (\ref{act}) provided the
dilaton and antisymmetric tensor fields are also expressed in
terms of the function $h(r)$:\ba
\phi_{\gamma}(r)&=&\frac{\gamma}{2}\log h(r),\label{deq}\\
H &=& -r^3 h' \Omega_3,  \label{3form} \ea where $\Omega_3\equiv
\sin^2 \!\theta \, \sin \varphi\, d\theta \wedge d\varphi \wedge
d\omega$ is the volume form on the three-sphere, and $dH=0$ except
at the location of the branes which act like sources. Finally,
$h(r)$ is a harmonic function satisfying
\begin{equation}
\mysquare  h = 0. \nn
\end{equation}

The general solution is therefore
\begin{equation}
h(z) = h_0+\frac{N}{r^2} ,
\label{heq}
\end{equation}
with $N$ and $h_0$ two integration
constants, which are both positive for $h(r)$ be positive. The
first one, $N\geq 0$, is interpreted as the total number of
five-branes, sitting at $r=0$, (integral of $dH$, vanishing
everywhere except for $r=0$).

If no five-branes are present, we recover flat-space with constant
dilaton and no antisymmetric tensor. This background is the target
space of an exact, albeit trivial, two-dimensional conformal sigma
model.

For $h_0=0$, the transverse geometry is an $S^3$ of radius
$L=\sqrt{N}$ with a covariantly constant antisymmetric tensor
(proportional to the three-sphere volume form) plus a linear
dilaton: \ba ds^2&=& -dt^2+d\vec{x}^2 +dy^2 + N\, d\Omega^2_3,
\label{nhg}\\
\phi_\gamma &=&\frac{\gamma}{2}\log N -\frac{\gamma y}{\sqrt{N}}
\label{lidi} \ea (we have introduced $y = \sqrt{N} \log r $, and
(\ref{nhg}) holds in the sigma-model frame).

Type-II strings in the geometry (\ref{nhg}), (\ref{lidi}) is an
exact $N=4$ superconformal theory \cite{CHS,kounnas,K93,AFK},
which implies the existence of $N=2$ space-time supersymmetry in
six dimensions (1/2 of the initial supersymmetry). From the
world-sheet point of view, this is an exactly conformal
two-dimensional sigma model if $\gamma = + 1 $ (NS). The target
space of the latter is the ten-dimensional manifold
$U(1)_Q{\times}SU(2)_{k}\times{M^6_{\vphantom k}}$. The last
factor is the flat six-dimensional longitudinal space--time,
$k=N-2$ is the level of the $SU(2)$ current algebra, and
$Q=-\gamma/\sqrt{k+2}$ is the background charge of the radial
transverse coordinate. This conformal field theory has been
extensively investigated in the past. Its spectrum is obtained by
appropriate combinations of $SU(2)_k$ and Liouville characters.
Discrete Liouville representations generate short $N=4$
multiplets, while continuous representations lead to long
(massive) multiplets. In the semiclassical limit, i.e. at large
$k$, one can trace back the origin of the various states: discrete
states are mostly confined in the vicinity of the brane, while
states from the continuous Liouville spectrum are delocalized in
the transverse directions. The latter are called {\it bulk
states}, while the former are the {\it brane states}, which are
those that survive in the decoupling limit, as excitations of the
little string theory. It should however be stressed once again
that separating the spectrum into ``bulk" and ``brane" states is
usually arbitrary, and at best valid in certain regimes only. In
fact, for an exact conformal field theory, even the geometry of
the target space itself is not clearly defined in all regimes.

The physical solution for the neutral five-brane \cite{CHS} has
non-vanishing $N$ and $h_0$. The corresponding background is
therefore asymptotically flat. This background does not correspond
to any (known) exact conformal field theory. However, it smoothly
connects two regions of space--time where the string propagation
leads to two different, and both exact conformal field theories
(at least for the NS branes). Other backgrounds, interpolating
between various exact-CFT target spaces, will be discussed in the
Secs. \ref{BFPRproj} and \ref{IKP}. Somehow, this seems to be a
natural feature of physical geometries.

\begin{figure}[htb]
\begin{center}
\epsfig{figure=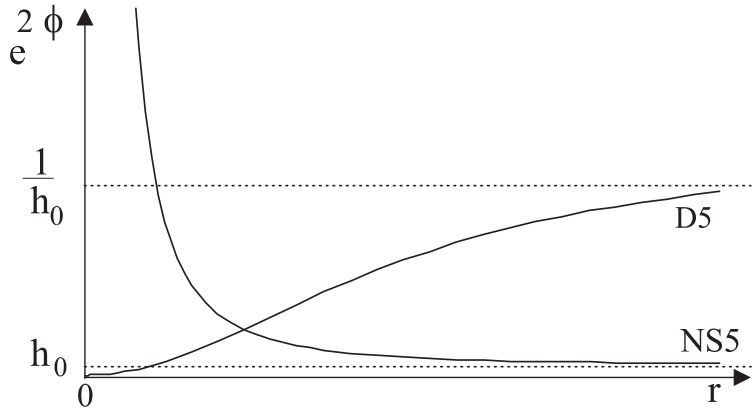, width=0.6\linewidth}
\caption{\label{divcoup} \small \small The string coupling of
solution (\ref{heq}) diverges at $r\to 0$ for the NS5-branes
($\gamma = + 1$). When the sources are D5-branes ($\gamma = -1$)
the divergence occurs at $r\to \infty$, provided $h_0 = 0$ (we
assume $h_0 < 1 $).}
\end{center}
\end{figure}

We now come to the strong-coupling problem. In the background
described by Eqs. (\ref{nhg}) and (\ref{lidi}), the string
coupling constant, $g_{\rm s} \equiv \exp \phi_\gamma= h^{\gamma/
2}$, becomes infinitely large at the location of the NS5-branes
($r=0$), while for the D5-brane background ($\gamma = -1$) the
same phenomenon occurs at $r \to +\infty$, i.e. in the asymptotic
region, far away from the sources (see Fig. \ref{divcoup}). In
these regions of space--time the {\it string perturbation breaks
down} and the very concept of worldsheet becomes questionable.

Although technically worrisome, the divergence of the string
coupling is not a pathology. Consider type-IIB string theory in
the background of NS5-branes. The $r\to 0$ limit corresponds to
the infra-red regime of the NS5's little string theory. In this
regime, the little strings are described by a $(1,1)$
six-dimensional super-Yang--Mills, which is free. Since the latter
is holographically dual to the original string theory in the
NS5-brane background, the string coupling must diverge where the
super-Yang--Mills coupling vanishes, i.e. at $r\to 0$. This
consistency check shows that the behavior of the string coupling
and in particular its divergence, is deeply related to the
structure of the theory itself. It can even be avoided within the
theory by performing an S-duality that switches to the D5-brane
background, where the $g_{ \rm s}\to 0$ in the vicinity of the
horizon. The caveats in this method are (\romannumeral1) the
absence of exact conformal field theory techniques for R--R
backgrounds, and (\romannumeral2) the treatment of the regions
where $g_{ \rm s} \sim 1$.

Similar considerations also apply  to the type-IIA NS5-brane
backgrounds. There, the little string theory in the infra-red is
the fixed point of a $(2,0)$ six-dimensional superconformal
theory. By holography, this is dual to a configuration of
M5-branes spread over a circle of vanishing radius. This is the
setup that should be worked out in order to capture the
strong-coupling regime of NS5-branes in type IIA. It necessitates
a complete understanding of M theory in the AdS$_7\times S^4$
background, which is presently out of reach.

In the absence of a satisfactory non-perturbative treatment of
string theory, the latter comments exhaust what can be done for
understanding the strong-coupling regime, {\it within} the above
five-brane backgrounds. One can however slightly {\it deviate}
from these backgrounds, or {\it embed} them in more general webs
of five-brane configurations. This can be helpful and will be the
subject of the following analysis, where our discussion will
mostly focus on Neveu--Schwartz backgrounds since those can
possibly lead to exact conformal field theories.

\section{Scanning the web of NS5 (or D5) branes}
\label{BFPRproj}

A rich variety of configurations can be reached by distributing
the five-branes in transverse subspaces. This generates new
backgrounds, which create remarkable webs of exact conformal field
theories, allow for new decoupling limits, and are ultimately
useful for understanding the strong-coupling problem.

The paradigm we will be following here is the distribution of
branes on a circle \cite{Sfetsos:1999xd}. There are several
degrees of approximation at which this system can be studied. We
will assume, as previously, a large number $N$ of branes, which
makes sensible the semi-classical treatment of the system. One can
consider the full geometry, or magnify the resolution in the
vicinity of the circle. In this region the geometry will be
sensitive to the nature of the distribution. If the $N$
five-branes are distributed on $\ell$ points carrying $n$ branes
each ($N=\ell \times n$), several regimes (decoupling limits) are
possible \cite{Sfetsos:1999xd, Giveon:1999zm, Giveon:1999zmp,BS,
Bakas:2000ax, Bakasfets, Kraus:1999hv,kounn,Giddings:2000zu,
Martelli:2002tu}. In particular, if the discrete structure remains
visible, the corresponding little string theory undergoes a Higgs
mechanism, where a $U(1)$ (translation along the circle) is broken
to $\mathbb{Z}_n$.

For the purposes of the present note, we will consider the
simplest situation, where we are close enough to the horizon to
avoid the asymptotic structure (such as e.g. $h_0$ terms in Eq.
(\ref{heq})), but at a distance where the discrete distribution is
not visible. In this regime, the transverse near-horizon metric
and  Kalb--Ramond field read:
\begin{equation}
ds^2=N\left(
d\rho^2+d\theta^2+\frac{\tan^2\theta \, d\psi ^2+\tanh ^2\rho \,
d\tau ^2}{1+\tanh^2\rho\tan ^2\theta}\right)\label{gcir}
\end{equation}
and
\begin{equation}
B= \frac{N}{1+\tanh^2\rho\, \tan^2\theta} d\tau \wedge d\psi.
\label{Bcir}
\end{equation}
We have used the transverse coordinates
$0\leq\rho<\infty$, $0\leq \theta\leq \pi/2$ and $0\leq
\psi,\tau\leq 2\pi$, in which the sources are located at $\rho =
0$ and $\theta=\pi/2$.

For NS5-branes, the dilaton field is given in
\begin{equation}
\e^{-2\phi}=\e^{-2\phi_0}\left(\cosh ^2\rho \, \cos^2\theta +
\sinh^2 \rho\, \sin^2 \theta \right), \label{dcir}
\end{equation}
and $g_{\rm s}$ diverges on the circle carrying the
sources\footnote{The constant $\phi_0$ is related to the coupling
$g_{\rm YM}$ of the little string theory in the usual manner:
$\exp {-2\phi_0}=\left({g_{\rm YM}^2U_0^2/N}\right)^{\gamma}$,
where $U_0$ is a constant.}.

Much like the simpler configuration of branes analyzed in Sec.
\ref{5brem}, the background (\ref{gcir})--(\ref{dcir}) turns out
to interpolate between various corners, where it coincides with
the target space of certain exact two-dimensional sigma models.
The above space can be  scanned as follows:

\underline{1. The $\rho=$ constant subspace}

The three-dimensional background obtained by setting $\rho=\rho_0$
is a squashed three-sphere. This is the target space of an
$SU(2)_k$ WZW model, marginally deformed with an exact $(1,1)$
operator, bilinear in some left and right $SU(2)_k$ currents:
$J\bar J$. In the four-dimensional transverse space under
consideration, this marginal deformation is {\it dynamical}, in
the sense that the corresponding parameter is promoted to genuine
coordinate, i.e. a two-dimensional dynamical field
\cite{Hassan:1992gi, Giveon:1993ph, Kiritsis:1993ju, DTC}.

\begin{figure}[htb]
\begin{center}
\epsfig{figure=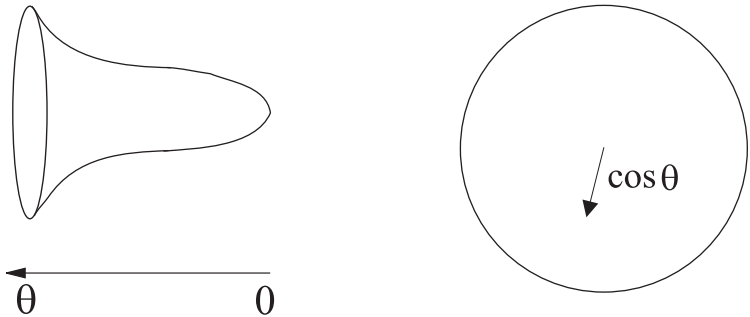, width=0.6\linewidth}
\caption{\label{bell} \small \small The bell geometry, with
angular and and radial coordinates $\psi$ and $\theta$. The
dilaton diverges at the boundary $\theta\to\pi/2$.}
\end{center}
\end{figure}

There are two limits of interest in the continuous line of
$SU(2)_k$ marginal deformations:

\noindent {\sl (a) $\rho\to 0$.} The limiting sigma model is in
this case $U(1)\times {SU(2)_k}\big/{U(1)}$. The first factor is a
line\footnote{It corresponds to the coordinate $\tau$ in the
metric (\ref{gcir}), appropriately rescaled, though.}, while the
target space of the coset factor is the {\it bell} geometry (see
Fig. \ref{bell}):
\begin{equation}
ds^2=N\left(
d\theta^2+\tan^2\theta \, d\psi ^2\right).\label{gbell}
\end{equation}
The string coupling
\begin{equation}
g_{\rm s}=\e^{\phi}=\frac{\e^{\phi_0}}{\cos\theta} \label{dbell}
\end{equation}
diverges on the boundary of the bell, which is
precisely the circle where the branes are distributed.

\noindent {\sl (b) $\rho\to \infty$.} The corresponding
three-dimensional target space is now the undeformed $S^3$
described by the $SU(2)_k$ WZW model.

\underline{2. The $\theta=$ constant subspace}

The interpretation of the three-dimensional space
$\theta=\theta_0$  as the target space of a conformal sigma model
is not clear except for two limiting values of $\theta_0$
\cite{Elitzur:cb, Forste:1994wp, Forste:2003km}.

\begin{figure}[htb]
\begin{center}
\epsfig{figure=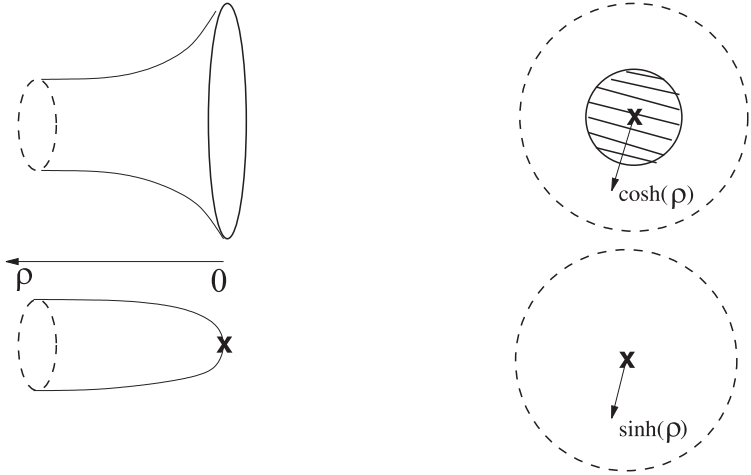, width=0.6\linewidth}
\caption{\label{trucig} \small \small The trumpet and cigar
geometries, with angular and radial coordinates $\psi$ or $\tau$
and $\rho$ respectively. The dilaton is finite everywhere on the
cigar. It diverges, however, at the boundary of the trumpet $\rho
\to 0$.}
\end{center}
\end{figure}

\noindent {\sl (a) $\theta= 0$.} The sigma model is now
$U(1)\times {SL(2, \mathbb{R})_k}\big/{U(1)}_{\rm axial}$ and the
coset factor describes the {\it cigar} geometry:
\begin{equation}
ds^2=N\left( d\rho^2+\tanh ^2\rho \, d\tau ^2\right)\label{gcig},
\end{equation}
with coupling
\begin{equation}
g_{\rm s}=\e^{\phi}=\frac{\e^{\phi_0}}{\cosh\rho} \label{dcig}
\end{equation}
everywhere finite (see Fig. \ref{trucig}).

\noindent {\sl (b) $\theta= \pi/2$.} The difference with the
previous case resides in the type of gauging which is used in the
coset. In the case at hand, this is a vector gauging with {\it
trumpet } geometry
\begin{equation}
ds^2=N\left( d\rho^2+{ \rm
cotanh } ^2\rho \, d\psi ^2\right)\label{gtru},
\end{equation}
and
coupling
\begin{equation}
g_{\rm s}=\e^{\phi}=\frac{\e^{\phi_0}}{\sinh\rho} \label{dtru}
\end{equation}
diverging at $\rho\to 0$.

The underlying two-dimensional conformal structure of the
background (\ref{gcir})--(\ref{dcir}) becomes somehow more
transparent after a chain of orbifold operations and T-dualities:
$\mathbb{Z}_N^\tau\to {\rm T}^\tau \to \mathbb{Z}_N^\psi\to {\rm
T}^\psi$. The resulting geometry,
\begin{equation}
ds^2=N\left( d\rho^2+d\theta^2+\tan^2\theta \, d\tau^2 +\tanh
^2\rho \, d\psi^2\right)\nonumber
\end{equation}
is the target space of the following conformal model:
\begin{equation}
\frac{SU(2)_k}{U(1)} \times \frac{SL(2,
\mathbb{R})_{k+4}}{U(1)_{\rm axial} }\label{Tdual}.
\end{equation}
The latter has $N=4$ superconformal invariance, and {\it no
strong-coupling problem} since the dilaton is constant; it has
been analyzed in \cite{K93,AFK,adk}. Moreover, the model
(\ref{Tdual}) is believed to be T-dual of the conformal model
$SU(2)_k \times U(1)_Q$ met in previous Sec. \ref{5brem} as
near-horizon geometry of $N$ coinciding NS5-branes. With this
starting point, several holographic issues of such a distribution
have been investigated in
\cite{Giveon:1999zm,Giveon:1999zmp,Gava:2002gv}.

Although T-duality in curved space is a subtle issue, and despite
the perturbative nature (in $\alpha'$) of many of the previous
results\footnote{Coset metrics such as (\ref{gbell}), (\ref{gcig})
or (\ref{gtru}) are expected to receive higher-order $\alpha'$
corrections.}, the whole picture appeals for further investigation
of the connections that might relate the various brane
distributions, and the role played by the underlying
two-dimensional sigma models, even if they appear as exactly
conformal only in some corners of the target space. This will
clearly help in understanding the issue of the strong-coupling
regime, which does not systematically occur.

\boldmath
\section{Spreading the NS5 (or D5) branes over an $S^3$}
\unboldmath\label{KKPR}

So far we have discussed the effect of distributing five-branes on
a circle in transverse space. The near-horizon geometries of the
backgrounds generated in that way exhibit remarkable properties,
and although they do not seem to be directly related to any known
conformal model, they possess smoothly connected corners
(subspaces) where we indeed recover CFT features. Exact conformal
description, reproducing the entire target space is possible only
after T-dualities.

Concerning the issue of the dilaton behavior, strong-coupling
regions still persist at the location of the sources. They
disappear only after some T-duality operations, which lead,
remarkably, to the target space of a CFT, ${SU(2)_k}\big/{U(1)}
\times {SL(2, \mathbb{R})_{k+4}}\big/{U(1)_{\rm
axial}}\times{M^6_{\vphantom k}}$, that turns out to be T-dual to
the original canonical five-brane near-horizon,
$U(1)_Q{\times}SU(2)_{k}\times{M^6_{\vphantom k}}$, studied in
Sec. \ref{5brem}. In some sense, this provides a way out to the
original strong-coupling problem, although not a satisfactory one:
in a curved non-compact background, T-duality may
\cite{Kiritsis:1993ju,Giveon:1994ph} or may not be an exact
symmetry\footnote{An NS5-brane with one longitudinal direction
wrapped on a circle is T-dual to flat space \cite{kkl}, although
we have serious reasons to believe that the dynamics in this case
is non-trivial. The Nappi--Witten pp-wave background
\cite{Nappi:1993ie}, which is also T-dual to flat space
\cite{kkl2}, is not equivalent to flat space or a standard
orbifold of it, and this can be asserted since its exact solution
is known \cite{kk, Elias}.}.

The picture in terms of five-branes on a circle may be therefore
an oversimplification. In this section we shall focus on another
option: distribute the  branes over a three-dimensional transverse
sphere. This configuration respects the transverse $SO(4) \simeq
SU(2)\times SU(2)$ symmetry, regulates truly the coupling, and
introduces a new kind of two-dimensional sigma-model whose target
space is a ``patchwork" of CFT target-space pieces.

The divergence of the ordinary Coulomb field can be avoided by
assuming a spherically symmetric distribution of charge over a
two-sphere centered at the original point-like charge. We can
similarly introduce a distribution of five-branes over the
transverse three-sphere \cite{Kiritsis:2002xr}, at some finite
radius, say $r=R$. This amounts in adding to the bulk action
(\ref{act}) a source term of the form:
\begin{equation}
S_{\rm five-brane}= -\frac{N\, T_5}{2\pi^2} \int d^{10}x \,
 \sin^2\!\theta \,
\sin \varphi \, \delta(r-R) \left(
\e^{-{\gamma\,\phi_\gamma}/2}\sqrt{-\hat{g}^{(6)}}+\tilde
C_{6}\right), \label{source}
\end{equation}
where $\tilde C_6$ is the dual of the two-index antisymmetric
tensor. Several remarks are in order here. In writing
(\ref{source}), we have chosen a gauge in which $(t,\vec{x})$ are
the world-volume coordinates of the five-branes. Thus, the induced
metric $\hat{g}^{(6)}_{\; \; ij}$ is just the reduction of the
background metric $g^{(10)}_{\; \; \mu \nu}$ ($\mu,\nu,\ldots \in
0,1,\ldots, 9$ and $i,j,\ldots \in 0,5,\ldots, 9$). All
five-branes are sitting at $r=R$, and are homogeneously
distributed over the $S^3$. Their density is normalized so that
the net number of five-branes be $N$.

The energy--momentum tensor of the source term (\ref{source}) is
\ba T^{\mu \nu}_{\rm five-brane}(x)&=&\frac{2}{\sqrt{-g^{(10)}}}
\frac{\delta S_{\rm five-brane}}{\delta g^{(10)}_{\; \; \mu \nu}(x)}\nn \\
&=&-\frac{N\, T_5}{2\pi^2}\sin^2\!\theta \, \sin \varphi \,
\delta(r-R)\e^{\alpha \gamma \phi_\gamma}\, \delta^{\mu}_i \,
\delta^{\nu}_j \, \hat{g}^{(6)\, ij}
\sqrt{\frac{\hat{g}^{(6)}}{g^{(10)}}}.\nn \ea This enables us to
write the full equations of motion resulting from action
(\ref{act}) plus (\ref{source}). Expressed in the sigma-model
frame, Eq. (\ref{source}) exhibits the following dilaton coupling:
$\exp -\frac{3+\gamma}{2}\phi_\gamma$. For $\gamma = + 1$ this is
indeed the coupling of an NS5-brane, while for $\gamma = - 1$ we
recover the D5-brane. Introducing the same ansatz as before for
the metric (Eq. (\ref{5met})), the dilaton and the three-index
tensor are given respectively by Eqs. (\ref{deq}) and
(\ref{3form}), in terms of $h(z)$, which  now solves
\begin{equation}
h(z)^{\frac{3}{4}}\e^{4z}\,\mysquare_{} h= -2 N \delta(z-Z).
\label{boxdelta}
\end{equation}
In writing the latter, we have introduced a new radial variable,
$z=\log r \in \mathbb{R}$, with $Z=\log R$ being the location of
the branes. We have also expressed\footnote{When $\alpha'$ is
restored (it has been set equal to one), the five-brane tensions
read $T_{\; 5}^{\rm NS}=\frac{2\pi^2 \alpha'}{\kappa_{10}^2}$ and
$T_{\, 5}^{\rm D}=\frac{1}{4 \pi^{3/2} \kappa_{10} \alpha'}$
\cite{bachas}. They turn out to be equal, once $\kappa_{10}$ is
expressed in terms of $\alpha'$: $2\kappa^2_{10}\equiv 16\pi
G^{\vphantom 2}_{10}= (2\pi)^7 {\alpha'}^4$, where $G^{\vphantom
2}_{10}$ is the ten-dimensional Newton's constant.} $T_5$ in terms
of $\kappa_{10}$. The result is independent of the nature of the
brane.

\begin{figure}[htb]
\begin{center}
\epsfig{figure=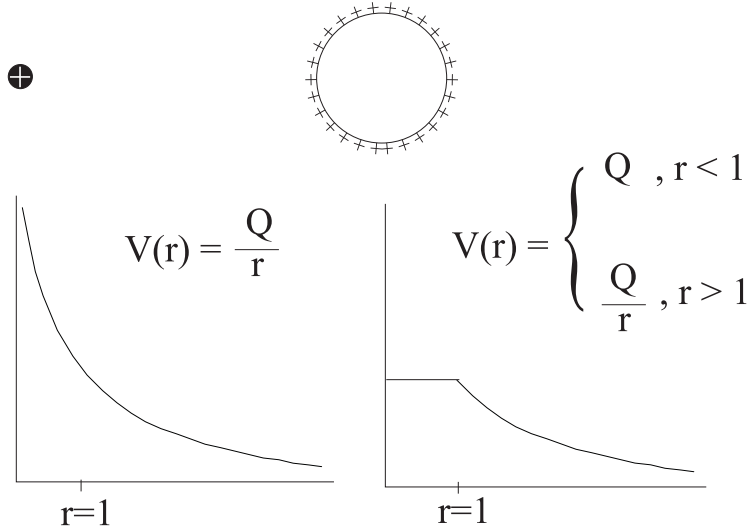, width=0.6\linewidth}
\caption{\label{ea} \small \small Electrostatic analogue of the
patched flat-space -- NS5-brane solution. The Coulomb potential of
a point-like charge $Q$ has a singularity at the origin, which is
resolved if the same charge is distributed over the surface of a
sphere (chosen here of unit radius). The Coulomb potential plays
the role of the dilaton field and the charge is $N$.}
\end{center}
\end{figure}

Replacing a point-like charge with a spherical distribution leads
to the same configuration outside the two-sphere, while the
electric field  vanishes inside (Gauss's law), avoiding thereby
the Coulomb divergence. This is depicted in Fig.  \ref{ea}. The
simplest solution to Eq. (\ref{boxdelta}), where we set for
simplicity $Z=0$ ($R=1$), is precisely an analogue of that
electrostatic example, as we have advertised previously:
\begin{equation}
h(z)=h_0 + N\e^{-(z+|z|)}. \label{electra}
\end{equation}
For $r>1$ ($z>0$) we recover (\ref{heq}), while for $0<r<1$
($z<0$) the space is flat since $h = h_0 +N$. Moving the brane
sources from $r=0$ to a uniform $S^3$ distribution at $r=1$
amounts therefore in excising a ball which contains the would-be
near-horizon geometry, and replacing it with a piece of flat
space. The price to pay for this matching is the introduction of
sources uniformly distributed over $S^3$ and localized at $z=0$.

The NS--NS or R--R flux (Eq. (\ref{3form})) now reads:
\begin{equation}
H = 2N \Theta(z) \Omega_3.  \label{3formstep}
\end{equation}
It vanishes inside the ball. Consequently $dH\sim \delta(z)$, and
its integral counts the total number of five-branes.

\begin{figure}[htb]
\begin{center}
\epsfig{figure=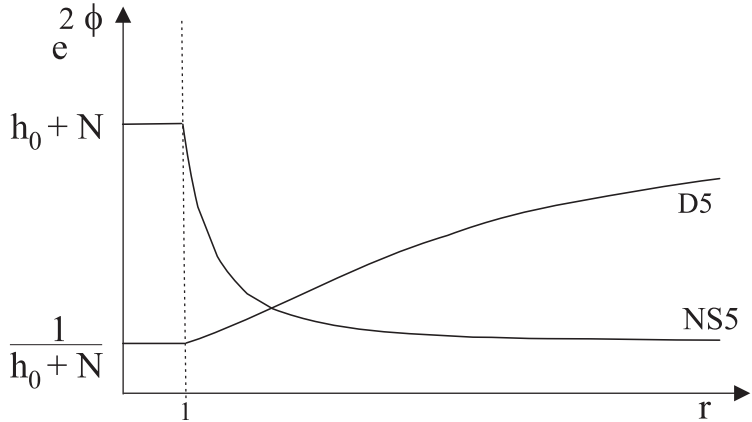, width=0.6\linewidth}
\caption{\label{coupling} \small \small The string coupling of
solution (\ref{electra}) is finite everywhere for the NS5-branes
($\gamma = + 1$) and its electrostatic analogue is given in Fig.
(\ref{ea}). However, it diverges when the sources are D5-branes
and $h_0 = 0$.}
\end{center}
\end{figure}

Concerning the dilaton field, Neveu--Schwarz and Dirichlet sources
lead to different pictures, according to Eqs. (\ref{deq}) and
(\ref{electra}). For NS5-branes, the excised ball removes
altogether the divergent-coupling region of the canonical neutral
five-brane, and replaces it with a constant one, $g_{\rm s}^2 =
h_0^{\vphantom 2} + N^{\vphantom 2}$. In the case of D5-branes,
the coupling inside the ball ($r<1$) becomes also constant,
$g_{\rm s}^2 = \left( h_0^{\vphantom 2} + N^{\vphantom
2}\right)^{-1}$. These results are summarized in Fig.
\ref{coupling}.

As we have already stressed, a remarkable situation is provided by
$h_0 = 0$. For negative $z$, the transverse space is flat, as for
generic $h_0$. For positive $z$, the geometry is that of a
three-sphere of radius $L=\sqrt{N}$ plus linear dilaton (see Eqs.
(\ref{nhg}) and (\ref{lidi})).

When the sources are of the Neveu--Schwartz type, both patches are
type-II string backgrounds described in terms of exact $N=4$
superconformal theories. In this case, the background is {\it free
of strong-coupling singularities.} This is to be compared to the
canonical situation where all branes are located at the origin. As
pointed out in Sec. \ref{5brem}, exact CFT interpretation is only
possible (\romannumeral1) in the vicinity of the origin, where the
near-horizon transverse geometry is precisely that of a
three-sphere of radius $L=\sqrt{N}$ plus linear dilaton, and
(\romannumeral2) in the asymptotic region, which is flat. By
moving the five-branes at finite $z$, we create a patchwork where
we sew finite (i.e. not only asymptotic) pieces of space--time,
each being a portion of some exact CFT target space. Although this
method can be easily generalized to more involved configurations
(see \cite{Kiritsis:2002xr}), it is not clear how this ``patchwork
CFT's" could be treated beyond the $\alpha'$ expansion.
Preliminary (perturbative) results about the spectrum and
holography are available though.

The case of Ramond--Ramond fields is more involved in all
respects. Firstly, there is no exact conformal model reproducing
such background fluxes. Secondly, spreading them over an $S^3$ at
finite distance might be interesting {\it per sei}, but does not
help in removing the strong-coupling region which, for $h_0=0$, is
at $z\to \infty$. Another electrostatic analogue is useful here in
order to resolve the divergence.

When a point-like charge is surrounded by a homogeneous spherical
shell of opposite charge, outside of the shell the potential is
constant whereas it is Coulomb inside. This is the screening
phenomenon. It can be transposed to the five-brane sources by
introducing negative-tension objects. These are orbifold planes in
the case of NS--NS backgrounds, and orientifold planes for R--R.
They cannot have fluctuations in a unitary theory because the
corresponding modes would be negative-norm.

In order to regulate the strong coupling in the configurations
where $N$ D5-branes are located at $z\to -\infty$ (i.e. at the
origin, $r=0$), we should introduce an equal number of orientifold
planes at some finite $z=Z$, so that the coupling remains constant
for $z\geq Z$. Whether many O5 planes could be included in a
theory is, to some extent, an open question. Obstructions do exist
for O9 planes, whereas there is in principle a small window for
accommodating a few O5's. The number of such planes might not be
allowed to exceed some finite value, though. Hence, the whole
scheme should be revisited, since the above geometrical
description assumes a large number of branes, especially when
those are distributed over an $S^3$ in a continuous fashion. If
such D5/O5 configurations are possible, they indeed regulate the
strong-coupling problem, without promoting the whole set-up to
some, even unconventional, conformal two-dimensional model.

On the contrary, there is no any constraint on the number of
orbifold planes that can be introduced for screening the charges
of NS5-branes. Although these planes are not necessary for solving
the strong-coupling problem, they allow to generate a large
variety of patchwork CFT backgrounds.

\boldmath
\section{Adding fundamental (or D) strings: null deformations of $SL(2,\mathbb{R})$}
\unboldmath \label{IKP}

There is another setup which offers a variety of possibilities for
supersymmetric string backgrounds, exact CFT realizations, new
decoupling limits and holographic pictures, and where the string
coupling is naturally regulated. This is the NS5/F1 or its S-dual
version (in type IIB), the D1/D5 \cite{Antoniadis:1989mn,
Maldacena:1998bw, Giveon:1998ns}. Those are $N=2$ backgrounds,
with $N=4$ enhancement in the near-horizon limit, where the
geometry is AdS$_3\times S^3\times T^4$.

The appearance of some near-horizon geometry is closely related to
a specific choice for the decoupling limit. The key observation is
here that one can introduce a tuning parameter that measures the
dilution of the strings and allows for finite and controllable
deformations of the AdS$_3$ factor \cite{Israel:2003}. The
remarkable fact is that those deformations turn out to be exactly
marginal deformations of the underlying conformal field theory,
namely the $SL(2,\mathbb{R})$ WZW model.

In order to be more specific, let us consider the D1/D5 picture.
The D5-branes extend over the coordinates $x\equiv x^5, x^6,
\ldots , x^9$, whereas the D1-branes are smeared along the
four-torus spanned by $x^6, \ldots , x^9$. The volume of this
torus is asymptotically $V= (2\pi)^4 \alpha^{\prime 2} v$ (we
restore $\alpha'$ in this chapter). With these conventions, in the
sigma-model frame, the supergravity solution at hand reads:
\begin{eqnarray}
 d\tilde s^2 & = & {1\over \sqrt{H_{1} H_{5}}}
\left(-dt^2 + dx^2 \right)+ \sqrt{H_{1}\over H_{5}} \sum_{\ell
=6}^9 (dx^i)^2 + \sqrt{H_{1}H_{5}} \left(dr^2 + r^2
d\Omega_{3}^{2} \right) , \label{sugrsol} \\
\mathrm{e}^{2\tilde \phi} & = & g_{\rm s}^2 {H_1 \over H_5},  \label{sugrsold}\\
H & = &  - \frac{1}{g_{\rm s}} d H_{1}^{-1}  \wedge dt \wedge dx +
2\alpha' N_5 \Omega_3\label{sugrsola}
\end{eqnarray}
($H$ is now the Ramond--Ramond three-form field strength) with
\begin{equation}
H_1 = 1 + \frac{g_{\rm s} \alpha' N_1}{vr^2} \ , \ \  H_5 = 1 +
\frac{g_{\rm s} \alpha' N_5}{r^2}.\nonumber
\end{equation}
The near-horizon ($r \to 0$) string coupling constant and the
ten-dimensional gravitational coupling constant are
\begin{equation}
g_{10}^2=g_{\rm s}^2 \frac{N_1}{vN_5} \  , \ \ 2 \kappa_{10}^{2} =
(2\pi )^7 \mathrm{e}^{2 \langle \tilde \phi \rangle }
\alpha^{\prime 4} .\nonumber
\end{equation}

The standard decoupling limit, which leads to the AdS$_3$/CFT$_2$
correspondence, is
\begin{equation}
\nonumber
\begin{array}{l}
\alpha'   \to   0, \\
U\equiv r/\alpha' \   \mathrm{fixed},\\
v   \   \mathrm{fixed}.
\end{array}
\end{equation}
In this limit, the holographic description is a two-dimensional
superconformal field theory living on the boundary of AdS$_3$ that
corresponds to the world-volume theory of the D1/D5 system
compactified on a $T^4$ whose volume is held fixed in Planck
units~\cite{Giveon:1998ns,deBoer:1998ip}.

In order to reach a decoupling limit that corresponds to the
near-horizon geometry for the D5-branes only, one has to consider
the limit:
\begin{equation}
\label{decoupl}
\begin{array}{l}
\alpha' \to 0, \\
U= r / \alpha' \ \mathrm{fixed}, \\
g_{\rm s} \alpha' \ \mathrm{fixed}, \\
\alpha^{\prime  2} v \ \mathrm{fixed}.
\end{array}
\end{equation}
The last condition is equivalent to keeping fixed the
six-dimensional string coupling constant:
\begin{equation}
g_{6}^2 = {g_{\rm s}^2 \over v}.\nonumber
\end{equation}
Since the gravitational coupling constant vanishes in this limit,
the world-volume theory decouples from the bulk. The geometrical
picture of the setup is the following: as $v \to \infty$, the
torus decompactifies and the density of D-strings diluted in the
world-volume of the D5-branes goes to zero.

The string coupling remains finite in this near-horizon limit,
while the asymptotic region is strongly coupled. A perturbative
description, valid everywhere is obtained by S-duality. The
supergravity solution (\ref{sugrsol}), (\ref{sugrsold}) in the
S-dual frame reads:
\begin{eqnarray}
ds^2 & = & \mathrm{e}^{-\tilde \phi} d\tilde s^2 = \frac{1}{g_{\rm
s}} \left\{ {1\over H_{1}} \left(-dt^2 + dx^2 \right)+ \sum_{\ell
=6}^9 (dx^{i})^2
 + \alpha^{\prime 2} H_5 \left(dU^2 + U^2 d\Omega_3^{ 2} \right)
 \right\}, \label{sugrsolS}\\
\mathrm{e}^{2\phi} & = & \frac{1}{g_{\rm s}^2}
\frac{H_5}{H_1}\label{sugrsolSd}
\end{eqnarray}
with (in the limit (\ref{decoupl}) under consideration)
\begin{equation}
H_1 = 1 + \frac{g_{\rm s} N_1}{\alpha' v U^2} \ , \ \ H_5 =
\frac{g_{\rm s} N_5}{\alpha' U^2}. \label{newH15}
\end{equation}
The expression (\ref{sugrsola}) for the antisymmetric tensor
remains unchanged but it stands now for a NS flux.

Finally, we define the new variables:
\begin{equation}
u= {1\over U}\ , \ \  X^\pm =
X\pm T = \frac{x \pm t} {g_{6}\sqrt{N_1 N_5}},\nonumber
\end{equation}
in which the  horizon is located at $u\to\infty$, while $u \sim 0$
corresponds to the asymptotic region. We also introduce the
following mass scale (in $\alpha'$ units):
\begin{equation}
M^2 = \frac{g_{\rm s} N_1}{\alpha' v}, \label{defparam}
\end{equation}
which measures the ``renormalized" dilution of D-strings. In these
coordinates, the solution (\ref{sugrsola}), (\ref{sugrsolS}) and
(\ref{sugrsolSd}), with (\ref{newH15}), reads:
\begin{eqnarray}
\frac{ds^2}{\alpha'}  &=&  N_5 \left\{ \frac{du^2}{u^2}+
\frac{dX^2-dT^2}{u^2 + 1/M^2} + d\Omega_{3}^{ 2}
\right\} + \frac{1}{\alpha' g_{\rm s}} \sum_{\ell =6}^9(dx^i)^2, \nonumber \\
\mathrm{e}^{2\phi}  &=&  \frac{1}{g_{10}^2} \frac{u^2}{u^2 + 1/M^2}, \label{solns} \\
{H\over  \alpha '} &=&   N_5 \left\{   \frac{2u}{\left(u^2 +
1/M^2\right)^2} du\wedge dT \wedge dX + 2\Omega_3
\right\}.\nonumber
\end{eqnarray}
This is the geometry of a {\it deformed AdS}$_3$ times an
$S^3\times T^4$. At highest possible D-string concentration ($M\to
\infty$), we recover the ordinary NS5/F1 near-horizon geometry,
AdS$_3\times S^3\times T^4$. In the opposite limit, namely for
infinitely diluted D-strings, the AdS$_3$ factor factorizes into
two light-cone coordinates plus a space direction with linear
dilaton. The remarkable fact is that not only these limiting
geometries are target spaces of conformal sigma models
(respectively $SL(2,\mathbb{R})_{k+4}\times SU(2)_k\times U(1)^4$
and $\mathbb{R}^{1,1}\times U(1)_Q\times SU(2)_k\times U(1)^4$,
$k=N_5-2$), but any geometry at finite $M$ originates from a
conformal two-dimensional theory. The latter turns out to be a
{\it marginal} deformation of $SL(2,\mathbb{R})$, driven by an
exact $(1,1)$ operator bilinear in null left and right currents.
Supersymmetry is $N=2$ along the line, with $N=4$ enhancement in
the two limits as well as at a subset of discrete values of $M$.

One can analyze the geometry (\ref{solns}) for arbitrary, finite
values of $M$. The asymptotic region ($u \to 0$) is not sensitive
to the presence of the fundamental strings. Hence, it is not
surprising that this region describes the near-horizon geometry of
the pure NS5-brane background, $\mathbb{R}^{1,1} \times U(1)_Q
\times SU(2)_k \times T^4$ {\it in its weakly coupled region --
i.e. away from the horizon}. In the opposite radial limit ($u\to
\infty$), one is  getting close the horizon and always feeling the
F1's: the background becomes effectively that of the NS5/F1
near-horizon: $SL(2,\mathbb{R})_{k+4}\times SU(2)_k \times T^4$,
with a finite constant dilaton. For any finite $u$ the geometry is
the deformed one, with a $u$-dependent dilaton, {\it bounded
everywhere}.

In some sense, we are here regulating the strong-coupling region
of the NS5-brane background by adding an appropriate condensate of
fundamental strings. This regularization is an alternative to the
one proposed in~\cite{Kiritsis:2002xr} and  described in Sec.
\ref{KKPR}; it avoids the spherical target-space wall of the
latter, and replaces it by a smooth transition, driven by a
marginal worldsheet deformation. Again, one can see how
interrelated are the issues of brane configurations, exact CFT
backgrounds and string-coupling behavior.

\section{Summary}

The near-horizon geometry of the canonical five-brane
configuration has been extensively investigated. It allows for a
decoupling limit, where the surviving degrees of freedom are the
brane excitations which exhibit a new dynamics, the little string
theory. Moreover, at least in the case of NS5-branes, the
perturbative string theory in this region is an exact conformal
field theory. Unfortunately, the perturbation theory is expected
to break down in this limit due to strong-coupling ambiguities.

In this note, we have tried to illustrate some pieces of
relationship among brane configurations, conformal models and the
strong-coupling problem, and reviewed the various methods that are
presently known for facing the latter, namely:

\noindent (\romannumeral1) Embed the original configuration in a
web of other configurations, related by T-dualities;

\noindent (\romannumeral2) Distribute the five-branes over a three
sphere in transverse space;

\noindent (\romannumeral3) Adding fundamental strings (or
D-strings in the S-dual picture), with a dilution parameter which
acts like a cut-off.

Further investigation is needed in all cases:  control the
T-dualities in curved spaces, incorporate the asymptotic regions
in a CFT framework, understand the decoupling limits in these more
exotic configurations, analyze the ``patchwork" conformal models
\dots

\vskip 0.56cm \centerline{\bf Acknowledgements} \vskip 0.25cm
\noindent We would like to thank our collaborators that have
contributed the works -- some yet unpublished -- from which this
note is inspired: C. Bachas, A. Fotopoulos, D. Isra\"el and S.
Ribault. We also thank the organizers of the meetings where these
ideas were presented.

This work was partially supported by European Union under the
contracts HPRN-CT-2000-00122, HPRN-CT-2000-00131,
HPRN-CT-2000-00148 and HPMF-CT-2002-01898.

\end{document}